\documentclass[journal=nalefd,manuscript=letter,layout = twocolumn]{achemso}
\setkeys{acs}{articletitle = false}
\usepackage{chemformula} 
\usepackage[T1]{fontenc} 
\usepackage{color}

\author{Ryota Akiyama}
\affiliation{Department of physics, The University of Tokyo, 7-3-1 Hongo, Bunkyo-ku, Tokyo, Japan}
\email{akiyama@surface.phys.s.u-tokyo.ac.jp}

\author{Ryo Ishikawa}
\affiliation{Institute of Materials, University of Tsukuba, 1-1-1 Tennoudai, Tsukuba, Ibaraki, Japan}
\altaffiliation{Current address: Future technology research laboratory, ULVAC, Inc., 2-1 Yamadaoka, Suita, Osaka, Japan}

\author{Kazuhiro Akutsu}
\affiliation{Neutron Science and Technology Center, Comprehensive Research Organization for Science and Society (CROSS), 162-1 Shirakata, Tokai, Ibaraki, Japan}

\author{Ryosuke Nakanishi}
\affiliation{Department of physics, The University of Tokyo, 7-3-1 Hongo, Bunkyo-ku, Tokyo, Japan}

\author{Yuta Tomohiro}
\affiliation{Institute of Materials, University of Tsukuba, 1-1-1 Tennoudai, Tsukuba, Ibaraki, Japan}

\author{Kazumi Watanabe}
\affiliation{Department of physics, The University of Tokyo, 7-3-1 Hongo, Bunkyo-ku, Tokyo, Japan}

\author{Kazuki Iida}
\affiliation{Neutron Science and Technology Center, Comprehensive Research Organization for Science and Society (CROSS), 162-1 Shirakata, Tokai, Ibaraki, Japan}

\author{Masanori Mitome}
\affiliation{Advanced Materials and Nanomaterials Laboratories, National Institute for Materials Science (NIMS), 1-1 Namiki, Tsukuba, Ibaraki, Japan}

\author{Shuji Hasegawa}
\affiliation{Department of physics, The University of Tokyo, 7-3-1 Hongo, Bunkyo-ku, Tokyo, Japan}

\author{Shinji Kuroda}
\affiliation{Institute of Materials, University of Tsukuba, 1-1-1 Tennoudai, Tsukuba, Ibaraki, Japan}

\title[An \textsf{achemso} demo]
  {Direct probe of ferromagnetic proximity effect at the interface in Fe/SnTe heterostructure by polarized neutron reflectometry}


\begin{document}

\begin{tocentry}
  \includegraphics[scale=0.582]{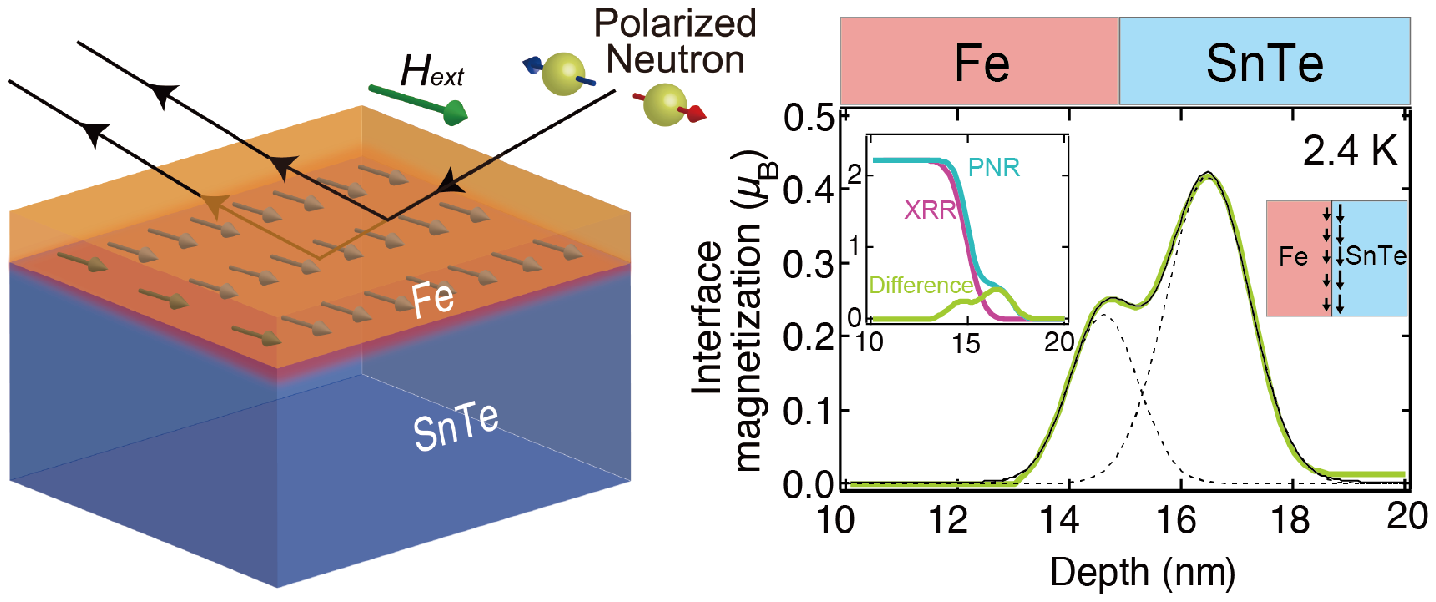}
\end{tocentry}

\begin{abstract}
Introducing magnetic order into a topological insulator (TI) system has been attracting much attention with an expectation of realizing exotic phenomena such as quantum anomalous Hall effect (QAHE) or axion insulator states. The magnetic proximity effect (MPE) is one of the promising schemes to induce the magnetic order on the surface of TI without introducing disorder accompanied by doping magnetic impurities in TI. In this study, we investigate the MPE at the interface of a heterostructure consisting of a topological crystalline insulator (TCI) SnTe and Fe by employing polarized neutron reflectometry.
The ferromagnetic order penetrates $\sim$ 3 nm deep into the SnTe layer from the interface with Fe, which persists up to room temperature. Our findings demonstrate that the interfacial magnetism is induced by the MPE on the surface of TCI preserving the coherent topological states, which is essential for the bulk-edge correspondence, without introducing disorder arising from a random distribution magnetic impurities. This opens up a way for realizing next generation electronics, spintronics, and quantum computational devices by making use of the characteristics of TCI.
\end{abstract}

\captionsetup[figure]{font=footnotesize}
\begin{figure*}[h]
 \begin{center}
  \includegraphics[scale=1.35]{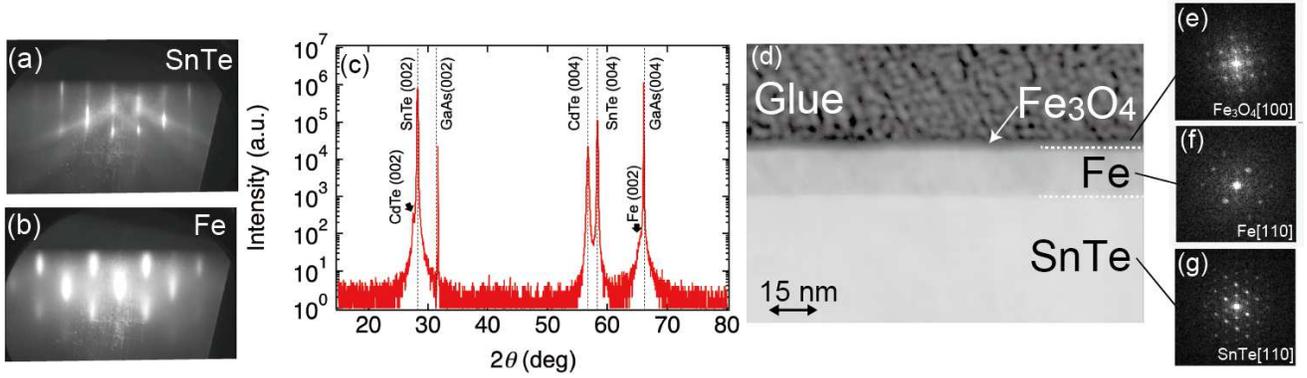}
  \caption{The results of structural characterizations on the heterostructure. (a) (b) RHEED patterns of just after the growth of (a) SnTe and (b) Fe layers with the incidence electron beam along [001] of the GaAs substrate. The streak pattern of Figure 1(a) indicates the atomically flat surface of the SnTe layer. The spotty pattern in Figure 1(b) indicates the epitaxial growth of the single-crystal Fe(001) layer, but with some roughness on the surface. (c) The result of 2$\theta$-$\omega$ XRD scan. The scan profile consists of the diffraction peaks from the (00$n$) planes of the Fe, SnTe, CdTe, GaAs layers, indicating the epitaxial growth of all these layers without any secondary phases. (d) Cross-sectional STEM image and (e)-(g) TED images of the (e) Fe$_{3}$O$_{4}$, (f) Fe, and (g) SnTe layers, respectively. The TED patterns of Figure 1(e)-(g) indicate that all the layers were grown epitaxially with in-plane orientational relations Fe$_{3}$O$_{4}$[100] $\parallel$ Fe[110] $\parallel$ SnTe[110]. The interface between the Fe and SnTe layers is abrupt without forming a mixing layer due to the diffusion of Fe into the SnTe layer. }
  \label{}
 \end{center}
\end{figure*}

Introducing magnetism into a topological insulator (TI) has now become a subject of intensive and extensive research related to topological physics and engineering. The breaking of time-reversal symmetry due to the onset of magnetism opens the gap (Dirac gap) in topological surface states (TSSs), which results in the realization of exotic phenomena such as quantum anomalous Hall effect (QAHE) or axion insulator states\cite{mogi,xiao} in response to the magnetic configuration with the Fermi level tuned within the Dirac gap. Especially, in the QAHE state, dissipation-less current flow is generated at the edge of a TI layer, which is expected to be utilized in novel energy-saving devices.

So far, the realization of the QAHE state has been reported in magnetic TI systems in which magnetic impurities such as Cr or V are doped in non-magnetic TI materials such as (Bi$_{x}$Sb$_{1-x}$)$_{2}$Te$_{3}$\cite{changsci,changnat,mogi2,checkelsky,kou,kandala,kou2,feng,grauer}, seemingly inspired from the past studies on diluted magnetic semiconductors. However, disorder introduced by random distribution of doped magnetic impurities should become a source of an inhomogeneity in the gap-opening, which is considered to be one of the origins of limiting the realization of the QAHE state at quite low temperatures.\cite{chen,lachman,Lee1316} In addition, the formation of precipitates of extrinsic phases or a degradation of crystallinity, which is inevitably induced by doping a large amount of magnetic impurities, would become another source of perturbation.
Alternatively, inducing magnetic order on the surface of TI due to the magnetic proximity effect (MPE) is one of the answers for introducing magnetism into a TI.\cite{katmis,li,li2,lee} In a heterostructure consisting of TI and ferromagnetic (FM) layers, the spatial separation between TI and FM allows us to evade the problems arising from the doping of magnetic impurities in TI. In particular, self-forming heterostructure consisting of an ordered stacking of TI and magnetic layers, such as MnBi$_{2}$Se(Te)$_{4}$, may be an ideal system for introducing magnetism in TI.\cite{hirahara,li2018,gong} Moreover, the MPE mediated by the Dirac surface state of the TI is expected to make the both of high mobility and spin manipulation compatible,\cite{li2,kim} which will be of great advantage in future devices.

Introducing magnetism in the TI surface states, either by bulk magnetic doping or by MPE in a TI/FM heterostructure, has been attempted for so-called $Z_{2}$ TIs so far, typically tetradymite materials such as Bi$_{2}$Se$_{3}$ and Bi$_{2}$Te$_{3}$. This is mainly because of relatively ease of the growth of these materials having layered structures and well-known physical properties. On the other hand, attempts to introduce magnetism for another class of topological materials, topological crystalline insulators (TCIs),\cite{tanaka,dziawa,hsieh2014} have rarely been demonstrated. TCIs are known to possess the TSS, similar to $Z_{2}$ TIs, but being protected by the structural symmetry of the crystal, instead of the time-reversal symmetry in $Z_{2}$ TIs. The topological properties in typical TCI materials, SnTe and related alloys, have experimentally been confirmed by observing the surface Dirac cone,\cite{tanaka,dziawa} weak antilocalization due to non-zero Berry phase $\pi$,\cite{assaf,akiyamaiop,akiyamanano} and Shubnikov-de Haas oscillations.\cite{safdar,dybko} Though the emergence of the TSSs in TCIs is quite similar to the case of $Z_{2}$ TIs, there are unique features characteristic of TCIs. In a typical case of SnTe having rock-salt (RS) structure, the metallic surface states with an even number of Dirac cones on high-symmetry crystal surfaces such as {001}, {110} and {111}, relying on the reflection symmetry with respect to the {110} mirror plane.\cite{fu,hsieh2012} When the surface orientation is changed in TCIs, arrangement of multi Dirac cones and the relation between an external magnetic field direction and a situation of the gap also change. Such ''controllable'' gap nature is inherent in TCIs and different from that in $Z_{2}$ TIs. Since most TIs have a layered structure (namely they can be cleaved in only one orientation naturally), it is difficult to prepare surfaces of various orientations; the Dirac cone arrangement is unique. However, with this tunable gap-opening among the multiple Dirac cones in TCIs, novel functionalities peculiar to TCIs are expected to be realized, such as electric tuning of the Dirac gap or control of the QAHE states among multiple quantized values. With an expectation to exploit these possibilities, we fabricate an epitaxially-grown heterostructure of Fe/SnTe with an atomically sharp interface and investigate the interface magnetism induced by the MPE by employing the polarized neutron reflectometry (PNR).

A thin film of single-crystal heterostructure of Fe/SnTe (15$\times$15$\times$0.5 mm$^{3}$ in size) was grown on a CdTe template \cite{ishikawa} by molecular beam epitaxy (MBE). SnTe was grown at a substrate temperature $T_{S}$ = 240$^\circ$C by supplying SnTe flux. Subsequently, $T_{S}$ was decreased down to 5$^\circ$C and Fe was deposited by supplying Fe flux. The detail methods are described in the methods section. As a result, the abrupt interface between Fe and SnTe layers was obtained as confirmed by the cross-sectional transmission electron microscope (TEM) observation (shown later).

The surface structure during the growth was monitored {\it in situ} by reflection high energy electron diffraction (RHEED). The observed RHEED patterns just after the growth of (a) SnTe and (b) Fe are shown in Figure 1, respectively. The streak pattern in Figure 1a indicates an atomically flat surface of the SnTe(001) layer. The RHEED pattern of the Fe layer (Figure 1b) indicates that single-crystalline Fe of bcc structure was epitaxially grown in the [001] orientation but the appearance of transmission-type diffraction spots suggests some roughness on the surface. This roughness is probably due to a strain induced by a slight lattice mismatch between SnTe(001) and Fe(001). There is an almost  commensurate relation between the bcc-Fe (001) lattice and the SnTe(001) lattice; five times of the unit cell of SnTe(001) ($a$ = 0.630 nm) almost matches (mismatch ratio < 1\%) to 11 times of that of Fe(001) ($a$ = 0.287 nm). 

To evaluate the crystallinity of the whole structure, we performed the 2$\theta$-$\omega$ X-ray diffraction (XRD) measurement. The scan profile shown in Figure 1c consists of diffractions from the Fe(001), SnTe(001), CdTe(001), and GaAs(001) planes, indicating that there is no second phase or precipitates. In order to check the abruptness of the interface between SnTe and Fe layers, which is critical to examine the MPE by comparing the density (X-ray reflectometry) and magnetic (neutron reflectometry) profiles, we observed the cross-sectional TEM. Figure 1d shows the cross-sectional scanning TEM (STEM) image and Figure 1e-g show transmission electron diffraction (TED) patterns of the respective layers. This result indicates that the heterostructure consists of the highly-ordered single-crystalline layers with a sharp interface between SnTe and Fe. The surface of the Fe layer was oxidized to form a thin layer of Fe$_{3}$O$_{4}$ of the inverse-spinel structure. 
The TED patterns of Figure 1e-g indicate that all the layers were grown epitaxially with in-plane orientational relations Fe$_{3}$O$_{4}$ [100] $\parallel$ Fe[110] $\parallel$ SnTe[110]; Fe$_{3}$O$_{4}$ was formed with 45$^\circ$ in-plane rotation with respect to the underlying Fe(001) layer.
\begin{figure}[!h]
 \begin{center}
  \includegraphics[scale=0.45]{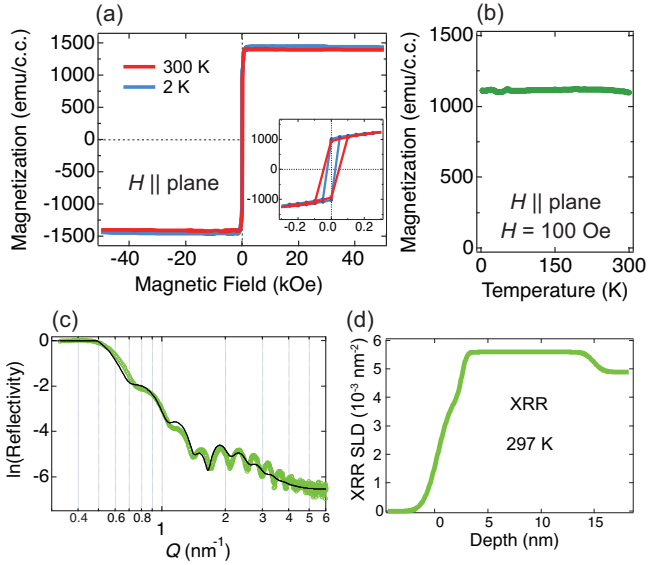}
  \caption{(a) Magnetic field dependences of in-plane magnetizations in the range of $H = \pm 50$ kOe at 2 K (blue) and 300 K (red). In the both curves, the magnetization are saturated at $\sim$ 1 kOe with a saturation magnetization of 1450 - 1400 emu/cc. The inset shows the same plot in the range of $H = \pm 300$ Oe in an enlarged scale. (b) Temperature dependence of magnetization under an external magnetic field of 100 Oe. (c) The result of XRR measurement at room temperature. The reflectivity is plotted as a function of momentum transfer $Q (= 4\pi/\lambda \sin\alpha)$ , where $\lambda$ and $\alpha$ are the wavelength (= 1.542 Å) and the incidence angle of X-ray, respectively. The black solid line represents the fitting curve. The estimated thickness of the Fe$_{3}$O$_{4}$ and Fe layers are 2.3 $\pm$ 0.96 nm and 12.4 $\pm$ 0.32 nm, respectively. The roughness at the interface of Fe/SnTe is estimated at 0.32 nm, which confirms the sharpness of the interface. (d) Depth profile of the scattering length density (SLD) deduced from the result of XRR measurement.}
  \label{}
 \end{center}
\end{figure}
The in-plane magnetization of the sample was measured using SQUID (MPMS-XL, Quantum Design). Figure 2a and b show the magnetic field ($H$) dependence of magnetization ($M$) at 2 K and 300 K, and the temperature ($T$) dependence of $M$ under a magnetic field of 100 Oe, both with external magnetic fields parallel to the plane. As shown in Figure 2a, the $M$-$H$ curves at 2 K and 300 K exhibit a sharp increase of magnetization around zero field, with a small hysteresis of coercive forces of 30 - 40 Oe, as shown in the inset. With increasing the magnetic field, the magnetization is saturated at $\sim$1 kOe with a saturation magnetization of 1400 - 1450 emu/cc, showing ferromagnetic behavior. In the $M$-$T$ curve shown in Figure 2b, the magnetization decreases only slightly with the increase of temperature from 2 K to 300 K. These results mainly reflect the magnetization in the Fe layer.
\begin{figure}[!h]
 \begin{center}
  \includegraphics[scale=0.48]{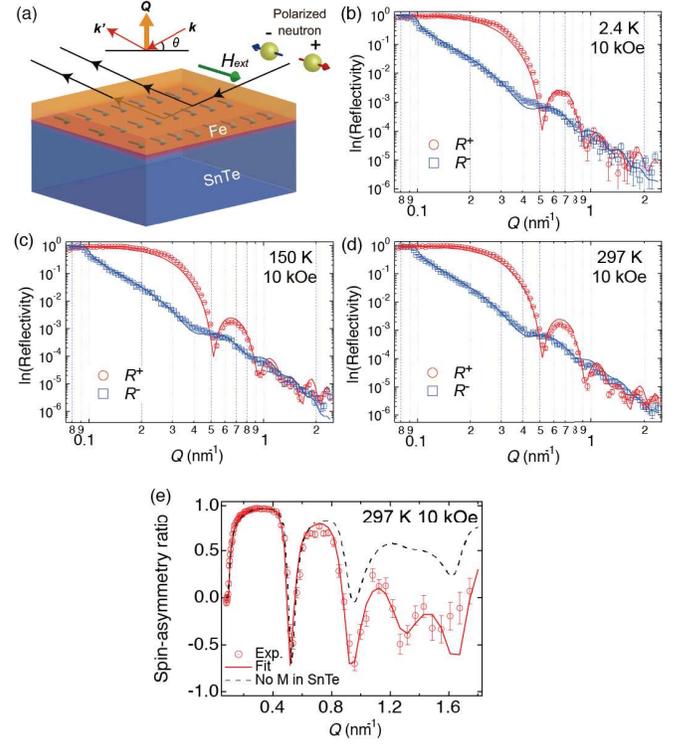}
  \caption{(a) Schematic picture of the PNR experiment. The collimated non-monochromatic polarized neutron beam is irradiated on the thin film with a grazing incidence angle $\theta$. The incident beam is reflected by atomic nuclei and the spins of unpaired electrons. The intensity of the reflected beam is recorded for the two opposite polarizations of neutron, $R^{+}$(red) and $R^{-}$(blue) with the time-of-flight method. Since the out-of-plane component of the magnetization parallel to the momentum transfer $Q$ (orange arrow) does not make a difference in the reflectance of neutron with the in-plane polarizations, it is not detectable in the PNR measurements. Only the in-plane component of magnetization can be detected. (b) - (d) The plot of $R^{+}$ and $R^{-}$ as a function of $Q$, derived from the measurements at 2.4 K, 150 K and 297 K under an in-plane magnetic field of $H =$ 10 kOe. The solid lines represent the fitting curves. (e) PNR spin-asymmetry (SA) ratio ($R^{+}$-$R^{-}$)/($R^{+}$+$R^{-}$), which is derived from the experimental and fitted reflectivities at 297 K under 10 kOe (shown in (d)). The black dashed curve represents the result of fitting without assuming the structure having the proximity layer (no magnetization is induced on the SnTe surface).}
  \label{}
 \end{center}
\end{figure}
To elucidate the structural depth profile of the heterostructure, we measured X-ray reflectometry (XRR) (SmartLab, Rigaku) at room temperature. The momentum transfer $Q$ (= 4$\pi$/$\lambda$ sin$\alpha$) dependence of reflectivity is shown in Figure 2c, where $\lambda$ and $\alpha$ are the wavelength (= 0.154 nm) and the incidence angle of X-ray, respectively. We fitted the reflectivity curve (green circles in the figure) to the theoretical one (black line) by using the Parratt's recurrence formula,\cite{Parratt} assuming the layer stacking of Fe$_{3}$O$_{4}$/Fe/SnTe. From the result of fitting, we derive the depth profile of the scattering length density (SLD), a measure of the scattering power of a material, which represents mainly electron density in the material in the case of X-ray scattering, as shown in Figure 2d.

The estimated thicknesses of the Fe$_{3}$O$_{4}$ and Fe layers are 2.3 $\pm$ 0.96 nm and 12.4 $\pm$ 0.32 nm, respectively. The thickness of SnTe (410 nm) was assumed to be infinity (backing layer) because it is thick enough for the reflection condition of the XRR measurement. The result of the XRR measurement only reflects the atom density distribution along the growth direction, without distinction of the chemical valence states. It is to be noted that the value of the standard deviation in the estimated thickness of the Fe layer is as small as 0.32 nm, which suggests that the interface between the Fe and SnTe layers is abrupt enough, with only a small degree of roughness.

To reveal the magnetism induced on the SnTe surface, PNR was measured at BL17 SHARAKU of the Materials and Life Science Experimental Facility (MLF) in the Japan Proton Accelerator Research Complex (J-PARC, Tokai, Japan).  An external magnetic field of 10 kOe was applied parallel to the sample plane. The schematic picture of the PNR experiment is shown in Figure 3a. In this configuration, the polarized neutron beam interacts with atomic nuclei and the spins of unpaired electrons, the latter of which corresponds to the magnetization of the material. The intensity of the reflected beam is recorded as a function of $Q$ for the two opposite polarizations of neutron, $R^{+}$ and $R^{-}$, where the superscript $+$ or $-$ corresponds to the neutron spin-parallel ($+$) or spin-antiparallel ($-$) to the direction of the applied magnetic field $H$, respectively. It should be noted that the out-of-plane component of magnetization is not detectable in the PNR measurements since the component of magnetization parallel to the momentum transfer vector $Q$ (perpendicular to the sample plane) does not make a difference in the reflectance of neutron with the in-plane polarizations; only the in-plane component of magnetization is observable.
Figure 3b-d show $Q$ dependences of the polarized reflectivities $R^{+}$ and $R^{-}$ under the in-plane magnetic field $H$ = 10 kOe at 2.4 K, 150 K and 297 K, respectively. In the measurements, we cooled the sample down to the respective temperatures under zero magnetic field and then applied the magnetic field of 10 kOe, the magnitude large enough to saturate the magnetization of the sample (Figure 2a). The reflectivity curves at all the temperatures can be fitted well by assuming the structure of Fe$_{3}$O$_{4}$/Fe/proximity layer/SnTe. The parameters derived from the fitting are shown in the Supporting Information. In Figure 3e, PNR spin-asymmetry (SA) ratio ($R^{+}$-$R^{-}$)/($R^{+}$+$R^{-}$), which is derived from the experimental and fitted curves of reflectivity at 297 K under 10 kOe, is shown as red circles and a red line, respectively. The black dashed curve in Figure 3e, derived from the fit assuming the structure without the proximity layer (no magnetization is induced on the SnTe surface), shows a large deviation from the experimental curve, especially at higher $Q$-region. Figure 4 shows the depth profiles of magnetic SLD (MSLD) obtained from the fitting of the neutron reflectivity curves at the respective temperatures. In the figure, the SLD profiles for the polarized reflectivities $R^{+}$ and $R^{-}$ (SLD${}_{R+}$, SLD${}_{R-}$) and the subtraction (SLD${}_{R+}$- SLD${}_{R-}$)/2 (MSLD) are plotted, together with the SLD obtained from the XRR measurement. Here, SLD${}_{R+(-)}$ and MSLD represent SLD of $R^{+(-)}$ and (SLD${}_{R+}$ - SLD${}_{R-}$)/2, respectively. The black dashed lines represent the structural boundaries determined from the depth profile of SLD by XRR, while the violet dashed line indicates the limit of extension of the magnetization induced at the interface. The obtained MSLD profiles clearly show that the magnetization penetrates into the SnTe layer beyond the structural boundary, with a step-like structure in the SnTe region at $\sim$ 1 - 3 nm away from the interface. This indicates ferromagnetic order is induced into the SnTe layer due to MPE.
\begin{figure}[!h]
 \begin{center}
  \includegraphics[scale=0.9]{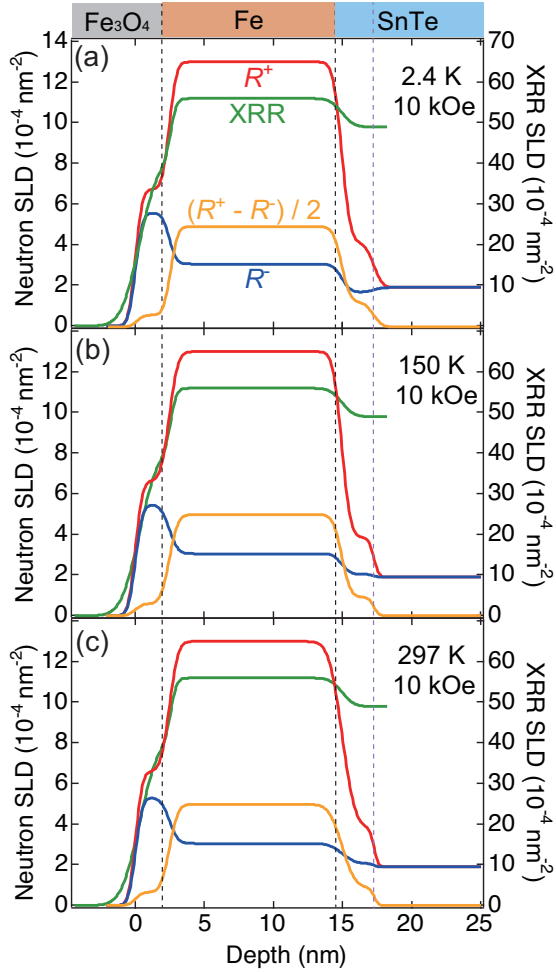}
  \caption{Depth profiles of SLD$_{R+}$ (red), SLD$_{R-}$ (blue) obtained by the fitting of the PNR curves of the polarized reflectivities $R^{+(-)}$ and MSLD (orange), the difference (SLD$_{R+}$ - SLD$_{R-}$)/2, at temperatures of (a) 2.4 K, (b) 150 K, and (c) 297 K, respectively. The profile of SLD obtained by XRR (green) is also plotted. The black dashed line represents the structural boundary while the violet dashed line indicates the limit of extension magnetization induced at the interface.}
  \label{}
 \end{center}
\end{figure}

To estimate the magnetism inside the SnTe region induced by the MPE, we focused on the difference of the depth profiles between SLD by XRR and MSLD in the region across the interface. SLD by XRR giving the depth profile of "material", which scales mainly with the electron density, and MSLD giving the depth profile of magnetization, the difference between them can be attributed to the interfacial magnetism induced by MPE. We, therefore, attempted to estimate the magnitude of the interfacial magnetism by taking a difference between the profile of SLD by XRR and that of MSLD in the following method. First, we estimated the profile of magnetic moment from that of MSLD, assuming that it is consistent with the reported value of 2.2$\mu_{B}$ at 2.4 K\cite{kittel} and decreases with temperature obeying the function using the Langevin function. Then, we normalized the SLD by XRR in such a way that the magnitude of SLD in the flat regions inside the Fe and SnTe layers coincides with that of MSLD, and then subtract the curve of the normalized SLD from the magnetic moment estimated from MSLD (see the insets of Figure 5). The subtracted curves shown in Figure 5, regarded as the profiles of the interfacial magnetism, have two peaks, which can be decomposed into two Gaussian peaks as represented by dashed curves in the figure. The temperature dependences of the fitting parameters such as the magnitude, peak position, and full width of half maximum (FWHM) of the two peaks are shown in Supporting Information. The most remarkable point is that peak 1 (a peak in the right) is located inside the SnTe layer, which indicates the ferromagnetic order is induced inside SnTe in the vicinity of the interface. With increasing temperature, the position of peak 1 hardly changes but the magnitude decreases and the width becomes narrower. Another peak, peak 2, is located inside the Fe layer at temperatures lower than 150 K but shifts into the SnTe layer at 300 K. 

Proximity-induced magnetism has been studied in $Z_{2}$ TI materials such as Bi$_{2}$Se$_{3}$ and Bi$_{2}$Te$_{3}$ so far; similar PNR experiments on heterostructures of $Z_{2}$ TI/ferromagnetic insulator revealed that the interfacial magnetism is induced on the surface of TI with a proximity depth of 1 - 2 nm in EuS/Bi$_{2}$Se$_{3}$\cite{katmis} and EuS/(Sb,V)$_{2}$Se$_{3}$\cite{li}. The penetration depth in SnTe revealed in the present study is comparable to these results in $Z_{2}$ TIs, or rather a little longer. This might be consistent with a theoretical prediction of stronger finite size effect in SnTe than Bi$_{2}$Se$_{3}$ \cite{Ozawa}. According to a calculation based on the tight-binding model, the gap opening induced by the hybridization of the top and bottom surfaces\cite{zhang2010crossover,li2010intrinsic,sakamoto} is more pronounced in SnTe compared to Bi$_{2}$Se$_{3}$; with increasing the layer thickness, the magnitude of the induced gap decreases more rapidly in Bi$_{2}$Se$_{3}$ than in SnTe\cite{Ozawa}. The different degree of the finite-size effect could be attributed to the difference in the crystal structure; in SnTe of a cubic structure, the bondings pointing to all the directions are equivalent, while in Bi$_{2}$Se$_{3}$ of a layered structure, the bondings are relatively weak between the layers. This difference in the bond strength between the cubic and layered structures would give an explanation for a different degree of penetration of the surface state and also the hybridization of the top and bottom surface states in a thin film\cite{gong2}. In the PNR experiments on a EuS/Bi$_{2}$Se$_{3}$ bilayer\cite{katmis}, it was observed that the magnetization was reduced in the EuS side near the interface. This reduction of the in-plane magnetization was ascribed to the canting of the Eu magnetic moments towards the perpendicular direction near the interface. On the contrary, in our case of Fe/SnTe, the magnetization is rather enhanced inside the Fe layer, as represented by the appearance of peak 2 in the profile of the interfacial magnetization shown in Figure 5. A possible explanation for this enhancement is that the Fe magnetization near the interface is enhanced due to the exchange interaction between $d$ electrons of Fe and Dirac electrons on the SnTe surface \cite{Jeong}. The reason of the different behaviors of the interfacial magnetization inside the FM layer remains to be clarified in further studies.

\begin{figure}[!h]
 \begin{center}
  \includegraphics[scale=0.84]{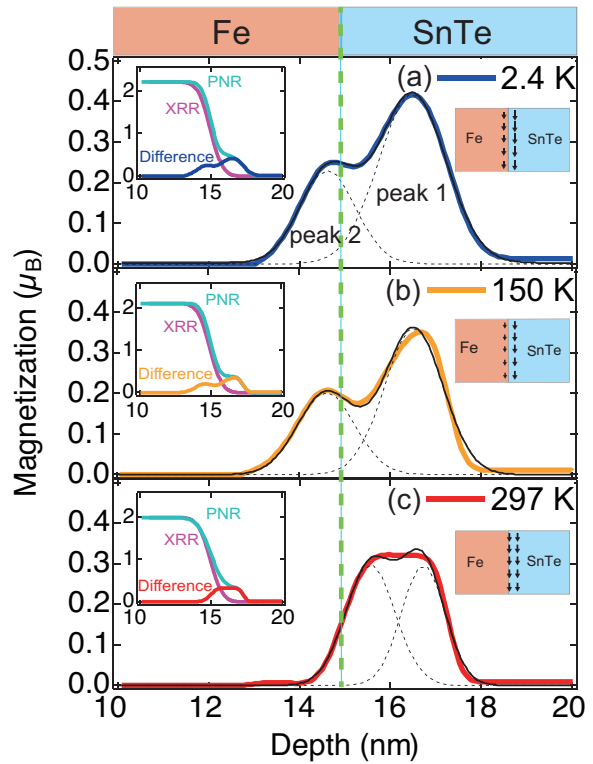}
  \caption{(a) - (c) Depth profiles of the interfacial magnetizations at (a) 2.4 K, (b) 150 K, and (c) 297 K, respectively. The profiles of the interfacial magnetizations indicated as "Difference" are obtained by subtracting the structural profiles of SLD by XRR (indicated as "XRR", purple curves) from the magnetization profiles of MSLD (indicated as "PNR", emerald curves). Here the green dashed line represents the structural boundary between the Fe and SnTe layers. The curves of the interfacial magnetization can be decomposed into two Gaussian peaks as represented by black dashed curves (main panel). The right insets show schematic pictures of the evolution of the in-plane magnetization under a magnetic field of 10 kOe. Black arrows indicate the interfacial magnetism. The proximity depth at 2.4 K is estimated at $\sim$ 3 nm. }
  \label{}
 \end{center}
\end{figure}

In summary, we investigated the MPE on the surface of SnTe, a typical material of TCI, by performing the PNR measurements on a single-crystalline heterostructure of Fe/SnTe grown by MBE. By comparing the structure profile derived from the XRR measurement and the magnetic profile from the PNR measurement, it was elucidated that the interfacial magnetism was induced inside SnTe due to the proximity effect. The induced magnetism penetrates into SnTe up to $\sim$ 3 nm from the interface, the extent of penetration being comparable to or rather greater than that observed in Bi$_{2}$Se$_{3}$, a typical material $Z_{2}$ TI. A larger proximity effect in SnTe revealed in this study paves a way for realizing dissipation-less spintronic devices by utilizing the QAHE or axion insulator state in this system.

\section{Methods}

A thin film of single-crystal heterostructure of Fe/SnTe (15$\times$15$\times$0.5 mm$^{3}$ in size) was grown on a CdTe template \cite{ishikawa} by molecular beam epitaxy (MBE).  A CdTe template was fabricated in an MBE chamber for the growth of II-VI compounds by growing a thick CdTe layer on a GaAs(001) substrate and successively capped with amorphous Te. Then, the CdTe template was introduced to another MBE chamber for the growth of IV-VI compounds. After introduced into the chamber, the template was first heated up to 240$^\circ$C for 4 min to desorb Te on the surface, and then SnTe was grown at a substrate temperature $T_{S}$ = 240$^\circ$C for 420 min by supplying SnTe flux with beam equivalent pressure (BEP) of 1.0 $\times10^{-7}$ Torr. At this $T_{S}$, it was confirmed that a single-crystal SnTe(001) layer was epitaxially grown. Subsequently, $T_{S}$ was decreased down to 5$^\circ$C and Fe was deposited for 37 min by supplying Fe flux with BEP of 1.0$\times10^{-8}$ Torr. The Fe layer was deposited at this low temperature in order to prevent the diffusion of Fe into the SnTe layer.\cite{ohya}

PNR measurements were performed at BL17 SHARAKU of MLF in J-PARC using a reflectometer with the polarized time-of-flight (TOF) mode.\cite{yamada2014}. An external magnetic field of 10 kOe was applied parallel to the sample plane using an electromagnet. The data reduction, normalization, and subtraction of the background were performed using a program installed in BL17 SHARAKU that was developed for the TOF and polarized neutron data. To fit the PNR profiles using the least-squares approach to minimize deviations, the Motofit program\cite{nelson2006} was used.The incident beam power of the proton accelerator was 150 kW for all the measurements, and a pulsed neutron beam was generated using a mercury target at 25 Hz. The wavelength ($\lambda$) range of the incident neutron beam was tuned to be approximately $\lambda$ =0.22 - 0.88 nm using disk choppers. The distance from the neutron source to the sample distance was 15.5 m, and the distance from the sample to a $^{3}$He gas point detector was 2.5 m. The incident angle for the sample varied from 0.3$^\circ$ to 2.7$^\circ$. The covered $Q$ range was $Q$ = 0.08 - 2.7 nm$^{-1}$. The polarized neutron beam of $\lambda$ = 0.22 - 0.88 nm was provided with a constant polarization efficiency (> 98.5$\%$) under a magnetic field of 10 kOe. A 13 mm-beam-footprint was maintained on the sample surfaces by using six kinds of incident slits, and the total exposure time for the measurements was 14 hours.

\section{Acknowledgement}
The XRR measurement was performed by courtesy of Prof. T. Suemasu in University of Tsukuba. The PNR experiment at MLF, J-PARC was carried out under the user program (Proposal No. 2017A0066). The TEM observation was carried out at the NIMS microstructural characterization platform and the sample was prepared for the TEM observation at the micro-fabrication platform of University of Tsukuba, both supported by "Nanotechnology Platform Program" of the Ministry of Education, Culture, Sports, Science and Technology (MEXT), Japan. This work was partially supported by a Grant-in-Aid for Scientific Research JP16H02108 and JP18H01857, a Grant-in-Aid for Young Scientists 26870086, a Grant-in-Aid for Challenging Exploratory Research JP18K18732, Innovative Areas ''Topological Materials Science'' JP16H00983 and JP15K21717, and ''Molecular Architectonics'' 25110010 from Japan Society for the Promotion of Science. One of the authors (SK) acknowledges the support from Center for Spintronics Research Network (CSRN), Osaka University, Japan.

\bibliography{REF} 
\bibliographystyle{unsrt} 

\end{document}